\documentclass[%
aip,
sd,%
amsmath,amssymb,
reprint,%
]{revtex4-1}
\usepackage[colorlinks=true,
linkcolor=red,
urlcolor=black,
citecolor=blue]{hyperref}
\usepackage{graphicx}
\usepackage{epstopdf}
\usepackage{dcolumn}
\usepackage{bm}

\begin{document}
\preprint{AIP/123-QED}
\title{Synchronization of moving oscillators in three dimensional space}
	\author{Soumen Majhi}
	\author{Dibakar Ghosh}\email{diba.ghosh@gmail.com}
	\affiliation{Physics and Applied Mathematics Unit, Indian Statistical Institute, Kolkata-700108, India}

	\date{\today}
	
	\begin{abstract}
		We investigate macroscopic behavior of a dynamical network consisting of a time-evolving wiring of interactions among a group of random walkers.  We assume that each walker (agent) has an oscillator and show that depending upon the nature of interaction, synchronization arises where each of the individual oscillators are allowed to move in such a random walk manner in a finite region of three dimensional space. Here the vision range of each oscillator decides the number of oscillators with which it interacts. The live  interaction between the oscillators is of intermediate type ( i.e., not local as well as not global) and may or may not be bidirectional.  We analytically derive density dependent threshold of coupling strength for synchronization using linear stability analysis and numerically verify the obtained analytical results. Additionally, we explore the concept of basin stability, a nonlinear measure based on volumes of basin of attractions, to investigate how stable the synchronous state is under large perturbations. 	The synchronization phenomenon is analyzed taking limit cycle and chaotic oscillators for wide ranges of parameters like interaction strength $k$ between the walkers, speed of movement $v$ and vision range $r$.
		
	\end{abstract}
	
	\maketitle
	\begin{quotation}
		{\bf  Recent researches attest to the fact that synchronization in time-varying complex network is very essential due to its possible applications in various fields. 		
			Particularly, in fields like mobile communications, robotics etc.  synchrony among moving and dynamic agents is certainly necessary. But only a few works on this subject have been reported so far. For some situations in coordinated motions of vehicles, flashing fireflies and swarming birds etc. interaction with any other agent lying in a specified zone may not always be possible. In this article, we try to resolve this issue by forming those zones in a slightly different manner and examine global synchrony in a network of mobile agents moving in a finite arena of three dimensional space, where an oscillator is assigned for each agent. In this context, both limit cycle and chaotic systems have been dealt with. The interaction between the oscillators depends on their proximity and the orientation of the movements as well. We elaborately study the influences of all the network parameters such as vision range, speed of motion and coupling strength both locally (using linear stability analysis) and non-locally (using basin stability analysis).}
	\end{quotation}

\section{Introduction}
In the last few decades, complex network theory has revealed that the type of  interaction between the nodes is very important in a network as far as the emergence of collective behavior such as synchronization is concerned.  Synchronization refers to a process wherein two (or many) systems adjust a given property of their motion to a common behavior due to  interaction between each other or to a forcing \cite{syper,sycha}. Recently, collective behavior of time-varying complex network has been an active research area due to its potential application in power transmission system \cite{pts}, consensus problem \cite{cons}, person-to-person communication \cite{mcom,mcom2,mcom3,mcom4}, to name only few.   In time-varying networks, connection topology of the network changes over time and emerging behaviors can be realized in such time-varying complex networks due to that varying interaction topology. Time-varying networks can be based on different types of interaction platforms, e.g., nodes of the network can be static in space with time-varying edges (coupling) between them and on the other hand, nodes can be moving in the space with the formation of the edges depending on the movement of the nodes.
	Time-varying network of the first type arises in functional brain networks discussed in \cite{fbn} where the outcomes confirmed that the brain connectivity patterns develop with time and frequency while preserving a small-world structure. Also in power transmission system \cite{pts}, where a random reconnection of links between nodes has been taken into account, in consensus problem \cite{cons} in which robustness to changes in multi-agent network topology due to node or link failures is analyzed and this structure has also been investigated in mobile communication \cite{mcom}.  These makes the analysis of the time-varying networks of the first type so important. Emergence of synchronization on these type of time-varying dynamical networks has been studied theoretically \cite{belykh,sinha,sychen1,sychen2} and experimentally \cite{danas}.  
\par On the other hand, there are so many examples of networks of movable physical agents that can interact with each other depending strictly on their proximity. For example, studies on group of animals, particularly where transition from disorder to order in mobile swarming desert locusts is reported in \cite{grani}. Coordinated motions of robots \cite{robot} where interactions within a group of mobile robots are investigated and proposed as a possible strategy for controlling the robots and also coordinated motions in vehicles \cite{vehicle} can be considered as such networks of movable physical entities in which regulated dynamics appear. Notably, there are also many examples where synchronization in moving oscillators play an essential role, e.g., in chemotaxis discussed in \cite{chemo} where the motion of each element is driven by the local gradient of the chemical density and the elements produce and consume this chemical in amounts that depend on its internal state. In wireless sensor networks \cite{sensor}, sensor nodes need to coordinate their operations and collaborate to achieve a complex sensing task, and also in mobile ad hoc networks \cite{adhoc}.
\par Therefore, systematic study on emerging behaviors in networks of moving oscillators deserve special attention. In this context, some early works include studying synchronization in networks of movable Kuramoto phase oscillators \cite{movkur} and also in mobile chaotic systems \cite{movch} where topology changes due to agent's motion. Spectral analysis of synchronization in such networks has also been done \cite{spect} and it has been shown that the spectral pattern differs between synchronization mechanisms. A dynamical network of moving oscillator is investigated to model the spread of infectious disease \cite{infect}. Such a network is also considered to show the effectiveness of nonlinear coupling over linear one in the context of synchrony of the network of moving oscillators \cite{punit}. Emergence of synchrony in a network of moving integrate-and-fire oscillators interacting with immediate neighbors upon firing time has also been analyzed \cite{if1,if2}.  Synchronization properties in ensemble of random walkers carrying internal clock in a periodic space under normal diffusive motion together with superdiffusion has been studied comprehensively \cite{rev1}. Influence of self-propulsion on the emergence of synchronization in dynamical networks of mobile oscillators is investigated \cite{rev2}  recently. Pattern formation in chemically interacting active rotors with self-propulsion \cite{rev3} needs to be addressed as well. 
\par Other notable works on synchronizability of a moving agent network includes \cite{wang,pori,chaos2016}. B. Kim et al.\cite{kim} investigated synchronous behavior of moving-agent network in three dimensional (3D) space based on the assumption of restricted interactions, where a certain number of fixed zones were preitemized and at any time, only those agents within these zones were supposed to interact with each other. This scheme may be related to the process of coordination in robotic networks.   
\par But, most of these works discussed above have been done by considering the movement of the agents solely in a finite region of two dimensional plane.  The movement of the agents in (3D) space could be more general platform of moving which is practically possible in many ways. In particular, the scenario of  moving oscillators in 3D space has a direct application in biological systems such as flashing fireflies which are moving in 3D space and interchange the light signal to search the potential mates \cite{firefl1,firefl2}.  More importantly, in the cases of coordinated motions in vehicles, flashing fireflies etc., for a particular agent, interaction with other agents lying inside a specified region like sphere or circle (in case of 3D or 2D spaces respectively) centered at that agent (proposed in most of the previous works), may not be possible. Rather, the connection may be formed depending on the direction of the movement and the agent may interact with only those, the agent can see (observe) and not with any others, staying on its back. For all these situations our algorithm (discussed in the next section) will be significant and crucial. 
\par  So, in this paper, we put the scheme of mobility in a general way and propose a finite (cubical) region of three dimensional space and study networks of limit cycle as well as chaotic oscillators. We discuss one fundamental emerging behavior in complex network, namely, synchronization. We numerically investigate the synchronization process for a class of moving oscillators including chaotic systems which are moving in 3D space.  Most of the earlier works \cite{movkur,movch,spect,infect,punit,if1,if2,wang} reported the synchronization behaviors in moving oscillators which are moving in two dimensional plane but little attention has been paid on synchronization in time-varying network where each oscillator is moving in 3D space. Here, in our scheme, the vision range decides the number of oscillators with which a particular oscillator interacts at an instant of time. We rigorously study the variation of order parameter for synchronization by changing three physical parameters, namely speed of movement of the agents $v$, vision range $r$  and interaction strength $k$.  We also derive the density dependent threshold of critical coupling strength for synchronization in the network using linear stability analysis. To explore the stability of the synchronous state against non-small perturbations, we use basin stability measure, which is a nonlocal and nonlinear measure of stability depending on the basin of attraction's volume.

\par The remaining part of this paper is organized as follows. In Sec. II, we discuss how the oscillators move in three dimensional space. In Sec. III, we present the mathematical form of the network of moving oscillators where network topology changes due to oscillator's motion. In Sec. IV, numerical results are shown for the moving oscillators network using limit cycle (Landau-Stuart) and chaotic (R\"{o}ssler) oscillators and we show how the level of synchrony depends upon the parameters like vision range, speed of movement and interaction strength.  Linear stability analysis of the synchronized state is discussed in Sec. V. In Sec. VI, we examine the stability of the synchronized state using basin stability approach.  Finally, we summarize our results in Sec. VII.

\section{ Random walk algorithm}
In this section, we demonstrate the scheme of random walk movement of oscillators in a finite region of three dimensional space. Initially, in physical space, $N$ number of  oscillators are randomly distributed in  $P\times Q\times R$ node mesh with a pseudo random number $R_i$ ( $0\le R_i\le1$) corresponding to the $i$-th  oscillator having position
coordinate $(\xi_i,\eta_i,\zeta_i)$, $i=1, 2, \cdots , N$. The algorithm makes all the oscillators to move either to right or left, to up or down, to front or back depending on the values of the pseudo random number associated with the oscillators. Two parameters are associated with the random walk in physical space, namely, the speed of movement of the oscillators $v$
and the vision range $r$.
\par The random walk movement of the oscillators in space is made by a simple change in the position coordinates $(\xi_i,\eta_i,\zeta_i)$ after each and every time iteration in the following way:
\begin{enumerate}
	\item[a)] If the pseudo-random number $R_i$ corresponding to the $i$-th oscillator lies in  $[0,\frac{1}{6})$ or $[\frac{1}{6},\frac{1}{3})$, then the movements along the +ve $x$-axis (the right) or -ve $x$-axis (the left) direction are given as :
	\begin{equation}
	\begin{array}{lcl}
	\xi_i(\tilde t+\delta \tilde t)=\xi_i(\tilde t)+v~\mbox{cos}\theta,\\
	\eta_i(\tilde t+\delta \tilde t)=\eta_i(\tilde t)+v~\mbox{sin}\theta,\\
	\zeta_i(\tilde t+\delta \tilde t)=\zeta_i(\tilde t)+v~ \mbox{sin}\theta,
	\end{array}
	\end{equation}
	corresponding to $\theta=0$ or $\theta=\pi$ respectively.\\Here $\delta \tilde t$ corresponds to the characteristic time for the movement of the agents.
	 Similarly, the movements along $y$-axes (to front or back) and $z$-axes (to up or down) corresponding to the cases of pseudo-random number $R_i$ respectively lying in $[\frac{1}{3},\frac{1}{2})$ or $[\frac{1}{2},\frac{2}{3})$ and $[\frac{2}{3},\frac{5}{6})$ or $[\frac{5}{6},1]$ is  defined.
	\item[b)] Updating time iteration to its next value, the described changes will be applied again for all the position coordinates $(\xi_i,\eta_i,\zeta_i)$ accordingly and so on.  For this, of course, $R_i$ varies with iteration. 
	\item[c)] Now according to the proposed algorithm without any suitable boundary conditions, the oscillators could be out of the physical space,  hence we impose periodic boundary conditions. Doing this, we make sure that all the oscillators remain in the physical space for all the course of time.
\end{enumerate} 
\par As already stated earlier, no oscillator interacts with all the other oscillators, rather they interact with only those that belong to a specified region (subregion) of that particular oscillator.
That means oscillators will interact with each other only when they are in a specified close proximity for a particular instant of time. Here in order to make it possible, inside the physical space in which oscillators are allowed to move, a much smaller region (subregion) of cubical structure is assigned to each and every oscillator.  We call the subregion for an oscillator as {\it vision size}.
\par
\begin{figure}[ht]
	\centerline{
		\includegraphics[scale=0.44]{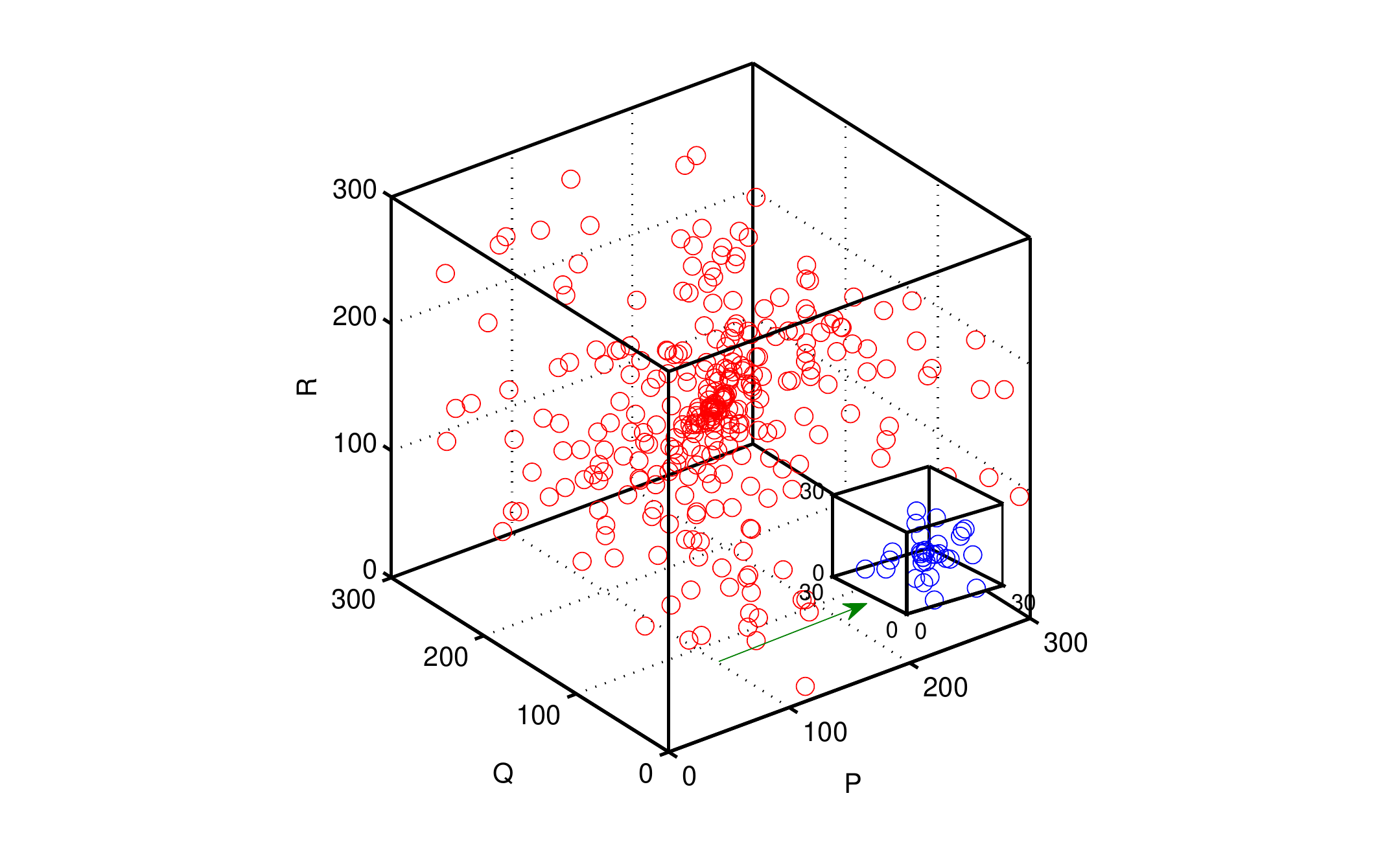}}
	\caption{  Moving oscillators (red circles) in three dimensional $P\times Q\times R$ node mesh. In particular, a cubical node mesh has been chosen with $P=Q=R=300$. The oscillators (blue circles) lying in a subregion of cubical structure is formed for a particular oscillator moving to its right, as indicated by the green arrow.}
	\label{3dmove}
\end{figure}

 Suppose at a particular instant of time $t$, an oscillator is moving to its right (say) which is the case of the corresponding pseudo-random number to be in $[0,\frac{1}{6})$, then a cube (the vision size) of volume $r^3$ is created to its right. Then the oscillator will interact  only with those  oscillators which lie in the created cube. But it does not interact with any oscillator to its left at that instant of time.
Similarly, if an oscillator is moving up, which is the case of the corresponding pseudo random number to be in $[\frac{2}{3},\frac{5}{6})$,  then a cube of volume $r^3$ is created in upward direction. Then the oscillator will interact with the oscillators which belong to its vision size but it will not interact with any oscillator in its downward direction.  Actually, the vision size (the cube) for a particular oscillator will be created in a direction along which the oscillator is moving. Creation of this type of vision size in the direction of motions is highly realistic in the cases of flashing fireflies, coordinated motions in vehicles etc. The same algorithm is applied to all the remaining cases. A graphical view of movement of oscillators and creation of vision size at a particular instant is shown in Fig.~\ref{3dmove}.

\section{Mathematical model of moving oscillators}
\par Under the above assumptions, we consider $N$ number of mobile agents or walkers which are moving in 3D space. Each agent $i~ (i=1,2,...,N)$ is a dynamical system whose state at time $t$ is described by $\dot{X_i}=F(X_i)$, where $X_i$ is a $m$-dimensional vector of dynamical variables of the $i$-th oscillator and $F(X_i)$ is a vector field. The interaction between the random walkers changes over time and the mathematical form of network of moving oscillators is
\begin{equation}
\dot X_i=F(X_i)-k\sum\limits_{j=1}^N g_{ij}(t)(EX_j),
\end{equation}
where $k$ is the control parameter accounting for the strength of the interaction between the random walkers. Here $G(t)=[g_{ij}(t)]_{N\times N}$ is a time-varying zero-row sum Laplacian matrix of order $N$ modeling the network connections at time $t$.  In particular, $g_{ij}(t)=-1$ if $j$-th oscillator lies in the vision size of $i$-th oscillator and otherwise zero.  Also, $g_{ii}(t)=n_i(t)$ where $n_i(t)$ is the number of oscillators which lie in the vision size of $i$-th oscillator at time $t$. The matrix $E$ is the $m\times m$ coupling matrix indicating which variables of one oscillator are coupled with which variables of the others. For example, coupling with only the first components, the form of $E$ is given by\\
$$ E = \left( \begin{array}{ccccc}
1 & 0 & 0 &\cdots & 0 \\
0 & 0 & 0 & \cdots & 0 \\
\cdot & \cdot & \cdot & \cdots & \cdot\\
0 & 0 & 0 & \cdots & 0
\end{array} \right).$$

\par Finally, the network of moving oscillators has three control parameters: the interaction strength between the random walkers $k$, the speed of movement of the agents $v$ and the vision range $r$ of the oscillators.  We fixed the parameters $P=Q=R=300$ and $N=100$ mainly, throughout this paper. We integrated the network equation (2) using fifth-order Runge-Kutta-Fehlberg algorithm with a time step size of $h=0.01$. The pseudo-random number is generated in FORTRAN 90. Initially, at time $t=0$, all the oscillators are randomly distributed in the physical space using pseudo-random number $R_i.$ In the following, we will explore the dynamics of the moving limit cycle and chaotic oscillators using the time-varying network (2), our main emphasis will be to identify the parameter range for three parameters $k, v$ and $r$ for synchronization.

\section{ RESULTS}
We first demonstrate how perfect synchrony emerges due to random walk movement in three dimensional space using coupled network (2). To do this, we take Landau-Stuart oscillator, when uncoupled, exhibits a stable limit cycle near supercritical Hopf bifurcation and has an unstable focus at origin. In the last two decades, 
Landau-Stuart oscillator has been used in the context of many emerging behaviors. So it will be interesting to explore our scheme using coupled Landau-Stuart
oscillators. We consider $N$ moving Landau-Stuart oscillators whose topologies change with the oscillators' motion in the form (2) with

\begin{equation}
F(X_i)=\left(
\begin{array}{c}
(1-{p_i}^2)x_i-\omega y_i\\
(1-{p_i}^2)y_i+\omega x_i\\
\end{array}
\right)
\end{equation}



where $p_i^2=x_i^2+y_i^2 ~~ (i=1, 2, ..., N)$, ~the coupling matrix is $E=[1~~ 0; 0~~ 0]$, and  $\omega$ is the intrinsic frequency of the individual Landau-Stuart oscillator. Only the $x$-components of Landau-Stuart oscillators are connected with each other via diffusive coupling.
\begin{figure}[ht]
	\centerline{
		\includegraphics[scale=0.80]{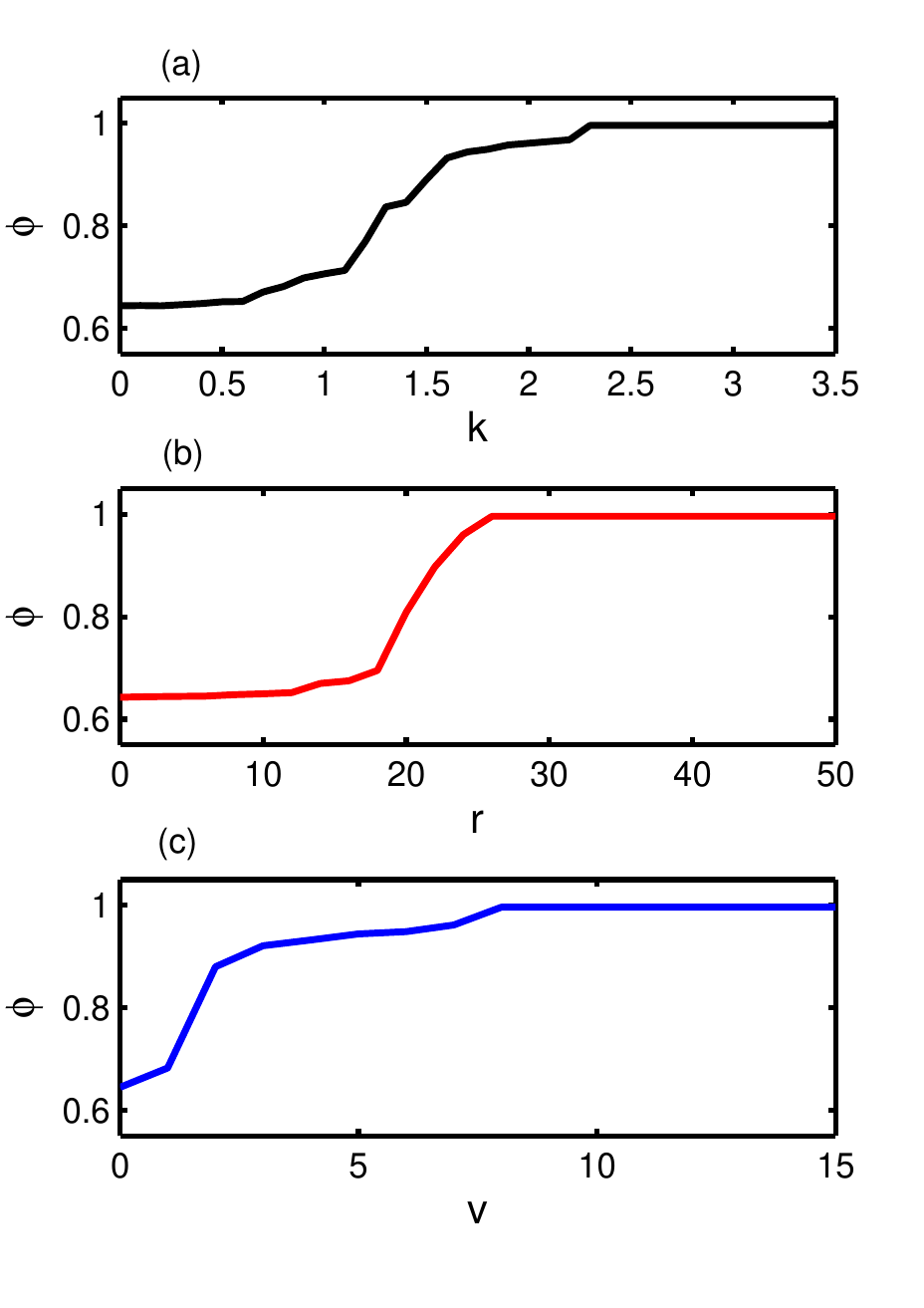}}
	\caption{ Moving Landau-Stuart oscillators: Variation of the order parameter $\phi$ by changing: $(a)$ the interaction strength $k$ for $v=12$ and $r=35$, $(b)$ vision range $r$ for $v=12$ and $k=3.0$, $(c)$ speed of movement $v$ for $k=3.0$ and $r=35$. Here $\omega=10.0$ and $N=100$.}
	\label{orderLS}
\end{figure}
\par  In order to quantify the level of synchrony, we calculate the order parameter $\phi$ as,
	\begin{equation}
	\phi=\left \langle \rho(t)\right \rangle_t,
	\end{equation}
	where $\rho(t)=|\frac{1}{N}\sum\limits_{j=1}^N e^{i \psi_j}| $, $i=\sqrt{-1}$, the phase $\psi_j(t)$ is calculated in $(x_j, y_j)$ plane as  $\psi_j(t)=\mbox{tan}^{-1}[\frac{y_j(t)}{x_j(t)}]$, $j=1,2,...,N$ and $\left \langle \cdots \right \rangle_t$ represents average over time. For complete synchronization, $\phi$ attains its value near unity while $\phi\ll 1$ when the oscillators are desynchronized. 
\par In Fig.~\ref{orderLS}, dependence of the order parameter $\phi$ \cite{initc} on different system parameters, namely interaction strength $k$, speed of movement $v$ and vision range $r$ are shown. Figure~\ref{orderLS}(a) shows how order parameter varies with respect to interaction strength $k$ where we fix other two parameters at $v=12$ and $r=35$. We observe that starting with a small value, $\phi$ starts to increase with increasing $k$ and finally gets saturated after reaching $\phi \simeq 1.0$, which is the maximum value it can achieve that indicates complete synchrony between the moving agents. Here the order parameter $\phi$ starts saturating after a certain value $k\simeq 2.3$ which means $k\simeq 2.3$ is sufficient to get perfect synchrony. Next we fix the interaction strength $k=3.0$ and speed of movement $v=12$ and vary the vision range of the agents. The variation of the order parameter $\phi$ with respect to vision range $r$ of the agents are shown in   Fig.~\ref{orderLS}(b).  As the vision range $r$ increases, we note that starting again with a small value, $\phi$  gets saturated after reaching $\phi \simeq 1.0$, which indicates complete synchrony among the moving oscillators. Thus as $r$ increases, the volume of the subregion and hence the possible number of oscillators in the subregion also increases, which makes the synchronization faster. In this case, the order parameter $\phi$ starts saturating after a certain value $r\simeq 26$ which means $r\simeq 26$ would be enough to get absolute synchrony. Finally, Fig.~\ref{orderLS}(c) shows change in the order parameter $\phi$ in terms of speed of movement $v$ for $r=35$ and $k=3.0$. As the speed $v$ increases, a type of long distance jump of the nodes occurs and we see that starting again with the small value, $\phi$ starts developing with increasing $v$ and saturation arises after reaching $\phi \simeq 1.0$  indicating complete synchronization between the moving oscillators. In this case the order parameter $\phi \simeq 1.0$ for $v\ge 8$.
\begin{figure}[ht]
	\centerline{
		\includegraphics[scale=0.50]{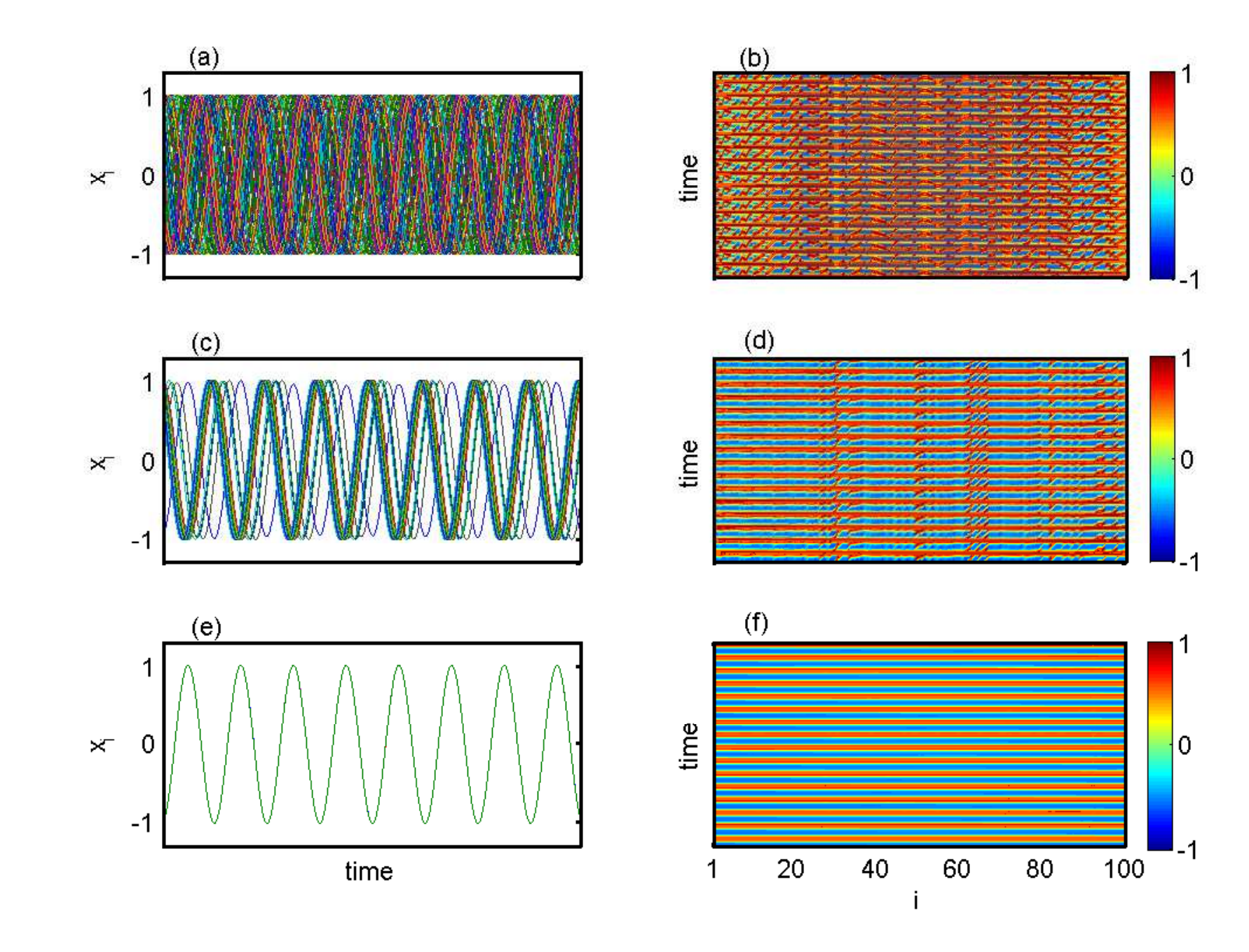}}
	\caption{ Time series (left column) and spatio-temporal (right column) plot of moving Landau-Stuart oscillators: (a, b) $k=0.2$, (c, d) $k=1.0$ and (e, f) $k=2.6$. The other parameters are $v=12$ and $r=35$.}
	\label{LScs}
\end{figure}
\par In order to understand the different synchronization regimes, we need to compare the time-series of moving agents for different physical system parameter values that lead the system to its perfect synchronization state. For this, we take three different values of $k$ from Fig.~\ref{orderLS}(a) where the other two parameters are fixed at $v=12$ and $r=35.$ Figures~\ref{LScs}(a) and \ref{LScs}(b) show the time series and spatio-temporal plot respectively for $x$-components of moving Landau-Stuart oscillators  for $k=0.2$. From these figures, it is noted that all the moving oscillators are in incoherent state (as expected, since the value of $\phi$ is much lower than unity). Next we consider the higher value of $k=1.0$, the time series and spatio-temporal plot of the oscillators are shown in  Figs.~\ref{LScs}(c) and \ref{LScs}(d) which clearly show that all the oscillators are not in complete synchronization state. But a coherence in the time series profiles can be seen for higher value of interaction strength $k=2.6$. At this point, the value of $\phi$ is almost unity and complete synchrony occurs. Finally, Figs.~\ref{LScs}(e) and \ref{LScs}(f) respectively illustrate the time series and spatio-temporal plot for complete synchrony state.

\begin{figure}[ht]
	\centerline{
		\includegraphics[scale=0.70]{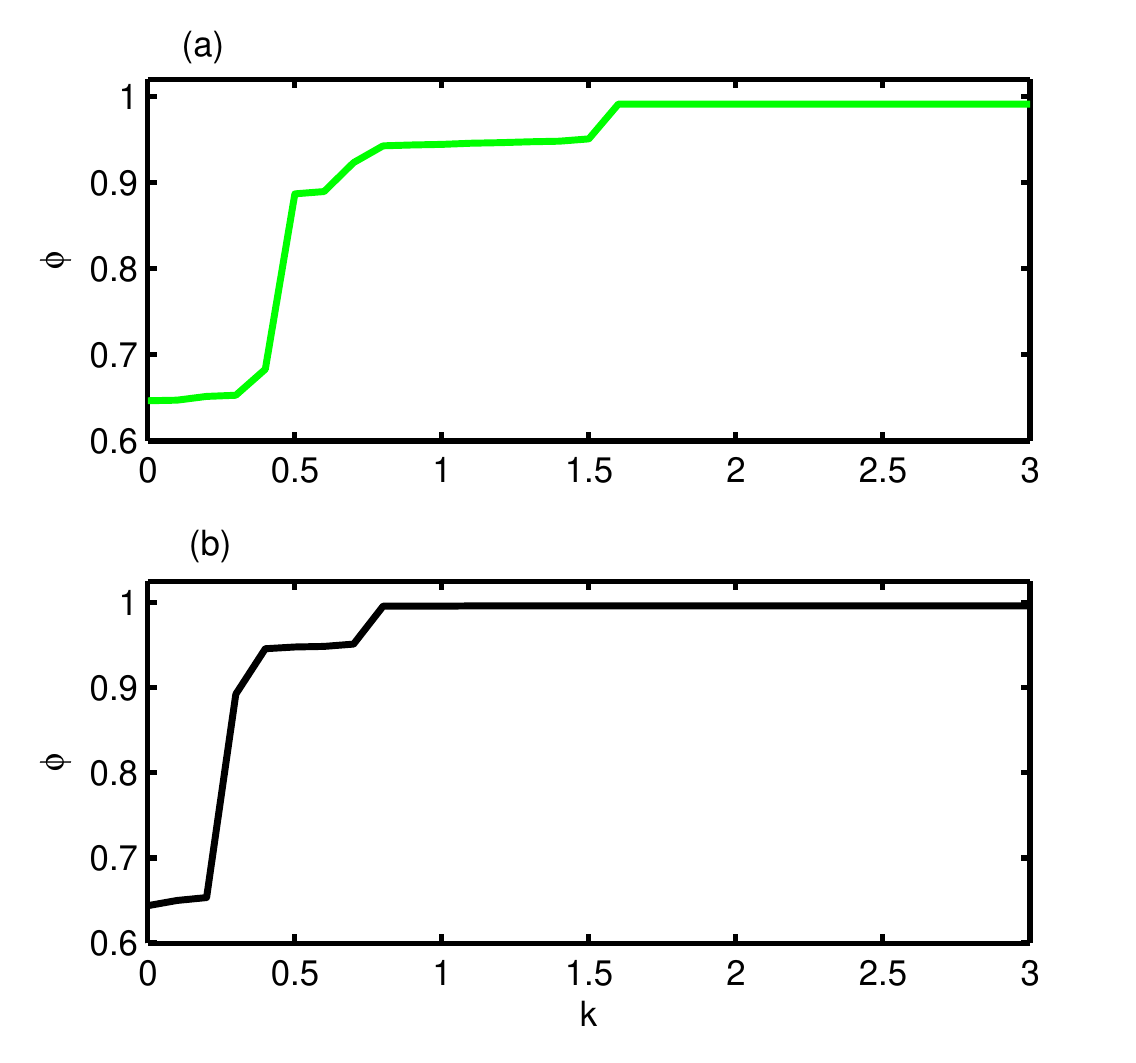}}
	\caption{ Moving R\"{o}ssler oscillators: Variation of the order parameter $\phi$ by changing the interaction strength $k$ with $v=8$ and $r=35$ for (a) $N=50$ and (b) $N=100$.}
	\label{orderRoss}
\end{figure}

\begin{figure}[ht]
	\centerline{
		\includegraphics[scale=0.70]{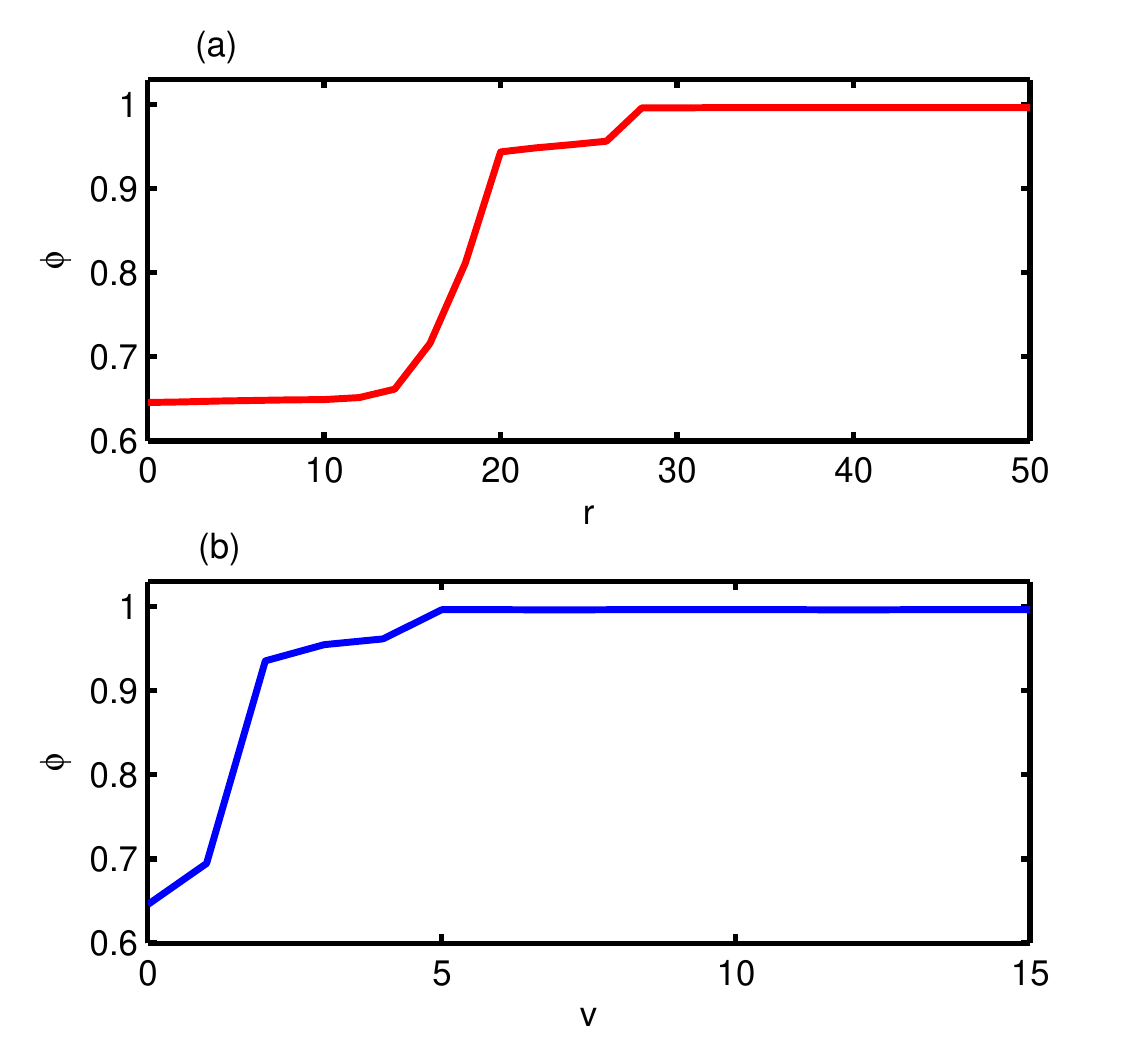}}
	\caption{ Moving R\"{o}ssler oscillators: Variation of the order parameter $\phi$ by changing the (a) vision range $r$ for $v=8$ and $k=2$; (b) speed of movement $v$ for $k=2$ and $r=35$. Here $N=100$.}
	\label{orderRoss2}
\end{figure}

Next we show that perfect synchrony appears due to random walk movement in space not only for oscillators of limit cycle type but also for oscillators that are chaotic in nature by taking moving R\"{o}ssler oscillators. We consider $N$ moving R\"{o}ssler oscillators in the form (2) where $F(X_i)$ has the form: \\
\begin{equation}
F(X_i)=\left(
\begin{array}{c}
-y_i-z_i\\
x_i+ay_i\\
b+z_i(x_i-c)\\
\end{array}
\right) \\
\end{equation}

with $X_i=(x_i,y_i,z_i)^T$. Each oscillator is in chaotic state for $a=b=0.1$, $c=14.0$. Only the $x$-components of the R\"{o}ssler oscillators are connected with each other, the coupling matrix $E$ is in the form $E=[1~~0~~0; 0~~0~~0; 0~~0~~0]$.

\begin{figure}[ht]
	\centerline{
		\includegraphics[scale=0.50]{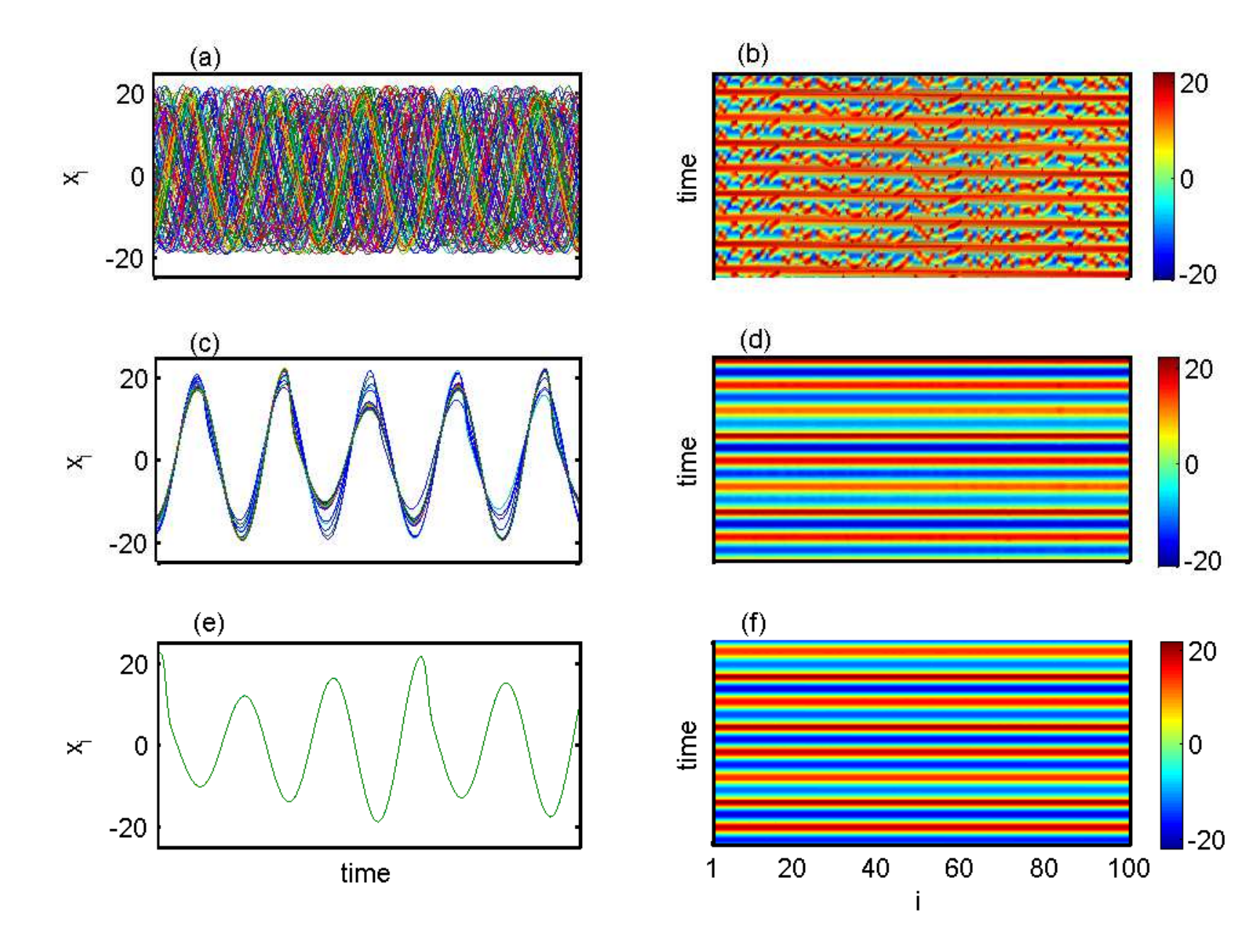}}
	\caption{ Time series (left column) and spatio-temporal (right column) plot of moving R\"{o}ssler oscillators for different interaction strengths: (a, b)  $k=0.1$; (c, d) $k=0.6$ and (e, f) $k=1.4$. Other parameters are $v=8$ and $r=35$. Here $N=100$.}
	\label{Ross_time}
\end{figure}
\par Now we analyze how the other physical parameters of the system, namely interaction strength $k$, speed of movement $v$ and vision range $r$ influence synchronization similar to the previous case. In Fig.~\ref{orderRoss}, variation of the order parameter $\phi$ for synchronization of moving R\"{o}ssler oscillators by varying $k$ for different $N$ are depicted. Figure~\ref{orderRoss}(a) shows how order parameter $\phi$ varies with respect to interaction strength $k$ with fixed values of $v=8$ and $r=35$ for $N=50$. Here the order parameter $\phi$ starts saturating with $\phi \simeq 1$ for a certain value $k\simeq 1.56$.  Figure~\ref{orderRoss}(b) shows how order parameter varies with respect to $k$ for $N=100$ with $v=8$, $r=35$ fixed as above. Similarly as before, the order parameter  $\phi$ starts developing with increasing $k$ and finally gets saturated for $k\simeq 0.78$ (where $\phi \simeq 1$) which means $k\simeq 0.78$ would be enough to get absolute synchrony. Figure~\ref{orderRoss2} characterizes the dependence of the order parameter $\phi$ on $r$ and $v$ with $N=100$ fixed. Figure~\ref{orderRoss2}(a) shows variation of order parameter $\phi$ with respect to vision range $r$ where the values of $v=8$ and $k=2$ are fixed. From this figure we see that for small values of $r$, all the moving oscillators are in desynchronized state as there are almost no links between the oscillators. As we increase this parameter $r$, we can see that synchronization of the moving oscillators appear for $r\geq 28$. Finally Fig.~\ref{orderRoss2}(b) shows change of order parameter in terms of speed of movement $v$ where $k=2$ and $r=35$.\\
\par  As already discussed, a saturation in the value of the order parameter $\phi$ can be seen no matter which physical system parameter, namely $k$, $v$ or $r$ is varied. So it is important to look at the time series profile to judge upto what extent it changes depending on the physical system parameters before and after $\phi$ gets saturated. To see how much the time series profile changes for $\phi$ before and after its saturation, we plot time series and spatio-temporal evolution for $x$-components of moving R\"{o}ssler oscillators for three different values of interaction strength $k$ at fixed values of $v=8$ and $r=35$ with $N=100$ in Fig.~\ref{Ross_time}. At lower value of $k=0.1$, incoherent states occur among the oscillators as in Fig.~\ref{Ross_time}(a) and corresponding space-time plot is shown in Fig.~\ref{Ross_time}(b). Figures~\ref{Ross_time}(c) and \ref{Ross_time}(d) illustrate the time-series and space-time plot for $k=0.6$, that show some of the oscillators are near to in-phase synchrony. By increasing the interaction strength to $k=1.4,$ complete synchrony occurs as the value of order parameter is $\phi\backsimeq 1.0$ there. Figures~\ref{Ross_time}(e) and \ref{Ross_time}(f) describe the time series and spatio-temporal plot respectively for $k=1.4$, which show that all the moving oscillators achieve complete synchrony.
\section{LINEAR STABILITY ANALYSIS}
 In this section, we analytically calculate the critical coupling strength for stable synchrony of moving oscillators. A time average Laplacian matrix is computed for the time-varying network using the procedure proposed by Stilwell et al. \cite{stilwell}. According to \cite{stilwell}, if the system of coupled oscillators given by \begin{equation}
	\dot X_i=F(X_i)-k\sum\limits_{j=1}^N \bar{g}_{ij}(EX_j),
	\end{equation} with fixed interaction topology (i.e., with fixed Laplacian matrix $\bar{G}=[\bar{g}_{ij}]$) admits a stable synchronization manifold and if there is a constant $\lambda$ such that $\frac{1}{\lambda} \int_{t}^{t+\lambda} G(\tau)d\tau=\bar{G}$, then there exists $\epsilon_0$ such that for all fixed $0<\epsilon <\epsilon_0,$ the system defined by \begin{equation}
	\dot X_i=F(X_i)-k\sum\limits_{j=1}^N g_{ij}(t/\epsilon)(EX_j),
	\end{equation} (i.e., coupled according to time-varying Laplacian $G(t/\epsilon)$) also carries a stable synchronization manifold. Thus, if the time average of the Laplacian matrix $G(t)$, given by $\bar{G}=\frac{1}{\lambda} \int_{t}^{t+\lambda} G(\tau)d\tau$ reflects synchrony of the system, then the time varying network will also support synchronization (provided that the switchings between the network configurations are fast).
	\par Here we note that in this work, the time scale of the agents' motion is considered much shorter than that of the oscillator dynamics. To find the time average Laplacian matrix $\bar{G}$, we start with the simplest case of $N=2$ for which four network configurations are possible at each time t. These network realizations are based on the following:
	\begin{enumerate}
		\item[i)] There is no interaction between the two nodes.
		\item[ii)] There is a unidirectional interaction from the node $2$ to the node $1$.
		\item[iii)] There is a unidirectional interaction  towards node $2$ from node $1$.
		\item[iv)] There exists bidirectional interaction between the nodes.
	\end{enumerate} 
	\par Then the time-average matrix $\bar{G}$ is given by
	\begin{equation}
	\bar{G}=p_{no} G_{no}+p_{u1} G_{u1}+p_{u2} G_{u2}+p_{bi} G_{bi},
	\end{equation}
	where $p_{no}$ is the probability that there is no interaction between the nodes and $p_{u1}$, $p_{u2}$, $p_{bi}$ are the probabilities corresponding to the cases of unidirectional interaction towards node $1$, towards node $2$ and bidirectional interaction among the nodes respectively. $G_{no}$, $G_{u1}$, $G_{u2}$ and $G_{bi}$ are the corresponding Laplacian matrices:\\
	$$ G_{no} = \left( \begin{array}{ccccc}
	0 & 0 \\
	0 & 0 \\
	\end{array} \right)\; \mbox{;}\; G_{u1} = \left( \begin{array}{ccccc}
	1 & -1 \\
	0 &  0 \\
	\end{array} \right)\; \mbox{;}\; G_{u2} = \left( \begin{array}{ccccc}
	0 & 0 \\
	-1 & 1 \\
	\end{array} \right)$$
	and
	$$G_{bi} = \left( \begin{array}{ccccc}
	1 & -1 \\
	-1 & 1 \\
	\end{array} \right).$$
	\par As $p_{u1}=p_{u2}=p_{u}$ (say), $\bar{G}$ becomes
	$\bar{G}=(p_u+p_{bi}) G_{bi}$. Now, $(p_u+p_{bi})$ being the probability of activation of a link between the nodes, $\bar{G}=p G_{bi}$, where $p=p_u+p_{bi}=\frac{r^3}{L^3}$ ($L^3=PQR$ is the volume of the considered physical space), under static node assumption and for moving node consideration $p\simeq\frac{r^3}{L^3}$ (small fluctuation may arise depending on $v$). Henceforth we will assume  $p=\frac{r^3}{L^3}$ and so $\bar{G}=p G_{bi}=\frac{r^3}{L^3} G_{bi}$. 
	\par In fact, it can be shown that the similar result holds for any number of nodes $N$ and $\bar{G}=p_g G_g$ with $p_g=\frac{r^3}{L^3}$ and $G_g$ is the $N\times N$ global (all-to-all) Laplacian matrix with bidirectional interactions. 
	 In fact, the above calculations can be repeated for any $N \geq 3$ (Confirmation using $N=3$ can be found in Appendix A) and finally we can conclude that
\begin{equation}
\begin{array}{lcl}
\bar{G}=\frac{\mbox{Volume of vision size}}{\mbox{Volume of physical space}} \;\;G_g
\end{array}
\end{equation}
 where
	$$ G_g = \left( \begin{array}{ccccc}
	N-1 & -1 & -1 &\cdots & -1 \\
	-1 & N-1 & -1 & \cdots & -1 \\
	\cdot & \cdot & \cdot & \cdots & \cdot\\
	-1 & -1 & -1 & \cdots & N-1
	\end{array} \right).$$ 
	\par 
	 As reported in \cite{msf1,msf2,msf3,msf4,msf5}, a
	block diagonalized variational equation of the form $\dot \eta_l=[DF-k\gamma_l DC]\eta_l$ represents the dynamics of the system around the synchronization manifold whenever the dynamics of the network is described by the equations
	\begin{equation}
	\dot X_i=F(X_i)-k\sum\limits_{j=1}^N \bar{g}_{ij}C(X_j), i=1,2,...,N,
	\end{equation}  where $C$ is the coupling function, $\gamma_l$ is the $l-$th eigen value of $\bar{G}=[\bar{g}_{ij}]_{N\times N}$, $l=1,2,...,N$ and $DF$, $DC$ are the
	Jacobian matrices corresponding to $F$ and $C$ around the state of synchrony. In terms of complex eigenvalues $\alpha+i\beta$ ($i=\sqrt{-1}$), the master stability equation becomes $\dot \xi=[DF-(\alpha+i\beta) DC]\xi$. Then the maximum Lyapunov exponent $\lambda_{max}$ of this equation serves the master stability function (MSF) as function of $\alpha$ and $\beta$. Thus the stbility of the synchronized state can be investigated from the eigenvalues $\gamma_l$ of $\bar{G}$ and looking at the negative sign of $\lambda_{max}$ at $k\gamma_l=(\alpha+i\beta)$.
	Now considering type $II$ MSF (for which the synchronization manifold is stable in the interval of type $[\beta_1,\beta_2]$), as in our case, $N$ eigenvalues of $\bar{G}$ are $\lambda_1=0$, $\lambda_j=p_g N, j=2,3,..., N$, there is a critical value of $k$, say $k_c$ that satisfies 
	\begin{equation}
		\begin{array}{lcl}
		k_c\lambda_j=\beta_1 
		\end{array}
		\end{equation}
		which implies
	\begin{equation}
	\begin{array}{lcl}
	k_c=\frac{\beta_1}{p_g N}=\frac{\beta_1 L^3}{r^3 N} 
	\end{array}
	\end{equation}
	with $\beta_1=0.1232$ (for coupled R\"{o}ssler systems).
	\par
\begin{figure}[ht]
		\centerline{
			\includegraphics[scale=0.75]{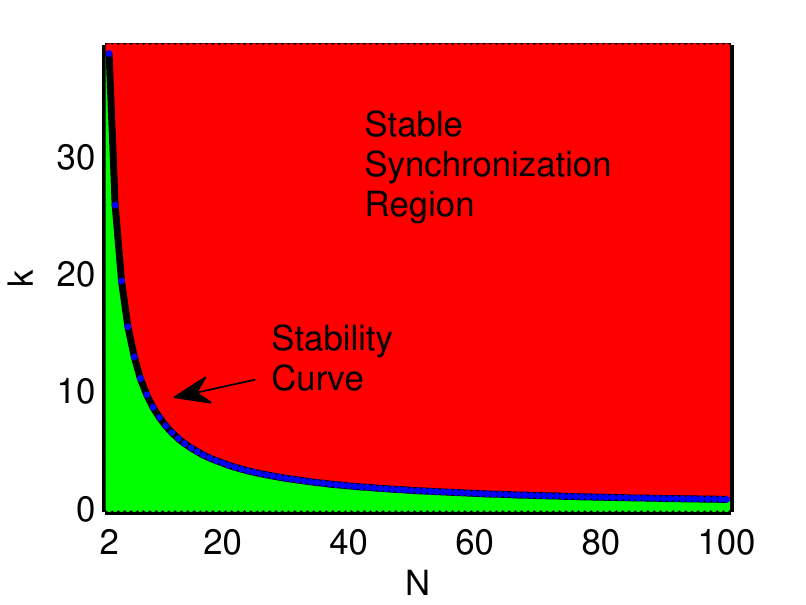}}
		\caption{ Moving R\"{o}ssler oscillators: Stable synchronization region (in red color) plotted in the $k-N$ parameter space. The arrow indicates the critical stability curve above which the stable synchrony appears. Here $r=35$ is taken.}
		\label{strgon}
	\end{figure}

 Clearly, according to the considered values of  $r=35$ and $L=P=Q=R=300$, for $N=50$, $k_c \simeq 1.56$ and for $N=100$, $k_c \simeq 0.78$, which are also the values of $k_c$ that we found for synchrony in Figs.~\ref{orderRoss}(a) and \ref{orderRoss}(b) respectively.  In fact, $k_c$ decreases with increasing $N$ according to the relation (12).  For complete understanding, the critical stability curve $k=\frac{\beta_1 L^3}{r^3 N}$ above which the network achieves stable synchrony is plotted in the $k-N$ parameter space in Fig.~\ref{strgon}.  Thus, the critical value $k_c$ for achieving synchronization decreases for increasing values of $N$ (i.e., the density of moving oscillators), which is quite expected as, if the number of nodes increases in the physical space of fixed volume, the possibility of interaction between nodes increases as well. Furthermore, pursuing the movement algorithm proposed in the article, the nodes move in the space at every iteration step, so that their neighbors (with which they interact) also get changed very frequently. And because of their movement in the whole space, effectively the nodes gain opportunity of interacting with so many (may be even all) other nodes during the entire course of time. Here these facts are inspected to be playing a deciding role in realizing global synchrony  at a lower coupling strength as the density increases. Hence for fixed interaction strength $k$, there will be critical value of population density $N$ for which synchrony in the network emerges.  This density-dependent threshold for achieving synchronization is related to that in quorum-sensing transition in indirectly coupled systems \cite{quo1,quo2,quo3}. This mechanism plays a decisive role in bacterial infection, biofilm formation and bioluminescence \cite{quo4}. However, in our algorithm the nodes are directly interacting with other nodes lying in the vision range. 
	
	\section{Basin stability analysis}
	The sections discussed above are made up of linearization-based study on synchronous state in network of moving oscillators. In other words, the results revealed that the synchronization state is stable for small perturbations near synchronous state and in some cases small perturbations could be infinitesimal. Such stability condition for  synchronization using linear stability analysis is necessary but not sufficient \cite{jost}.  Recently, Menck et al. \cite{bs} proposed a non-local stability concept to quantify how stable a state is against non-small perturbations in terms of basin stability (BS) which reflects the fact that both linear stability and BS approaches should be dealt with in order to check out the stability of the synchronized state. So here, we are concerned with the basin stability analysis of the synchronized and desynchronized state of the moving oscillators' network. BS of the synchronization state gives information about the relative volume of basin of attraction that approach to the synchronous state.  In contrast to linear stability, basin stability is a nonlinear as well as a nonlocal measure that determines how stable a state is while facing non-small perturbations. In this case, basin's volume quantifies the state's stability. 
	\par To calculate the basin stability for R\"{o}ssler system, we use the following procedure: We randomly choose sufficiently large number of initial conditions within the volume $\{-20\le x_i\le 20, -20\le y_i\le 20, 0\le z_i\le 40\}, i=1,2,...,N$. For each fixed value of network parameter (e.g. coupling strength $k$), we integrate the network equation (2) together with (5) for $P(=5000)$ number of random initial conditions from the prescribed volume and identify $P_s$ number of initial conditions for which $\phi\simeq1$ \cite{tol}. Then the basin stability at that network parameter is
	\begin{equation}
	\begin{array}{lcl}    
	BS=\frac{P_s}{P}
	\end{array}
	\end{equation}
	The value of $BS$ changes in the range from 0 to 1. $BS=1$ means that the network exhibits stable synchronization state for large number of initial conditions.
	\begin{figure}[ht]
		\centerline{
			\includegraphics[scale=0.75]{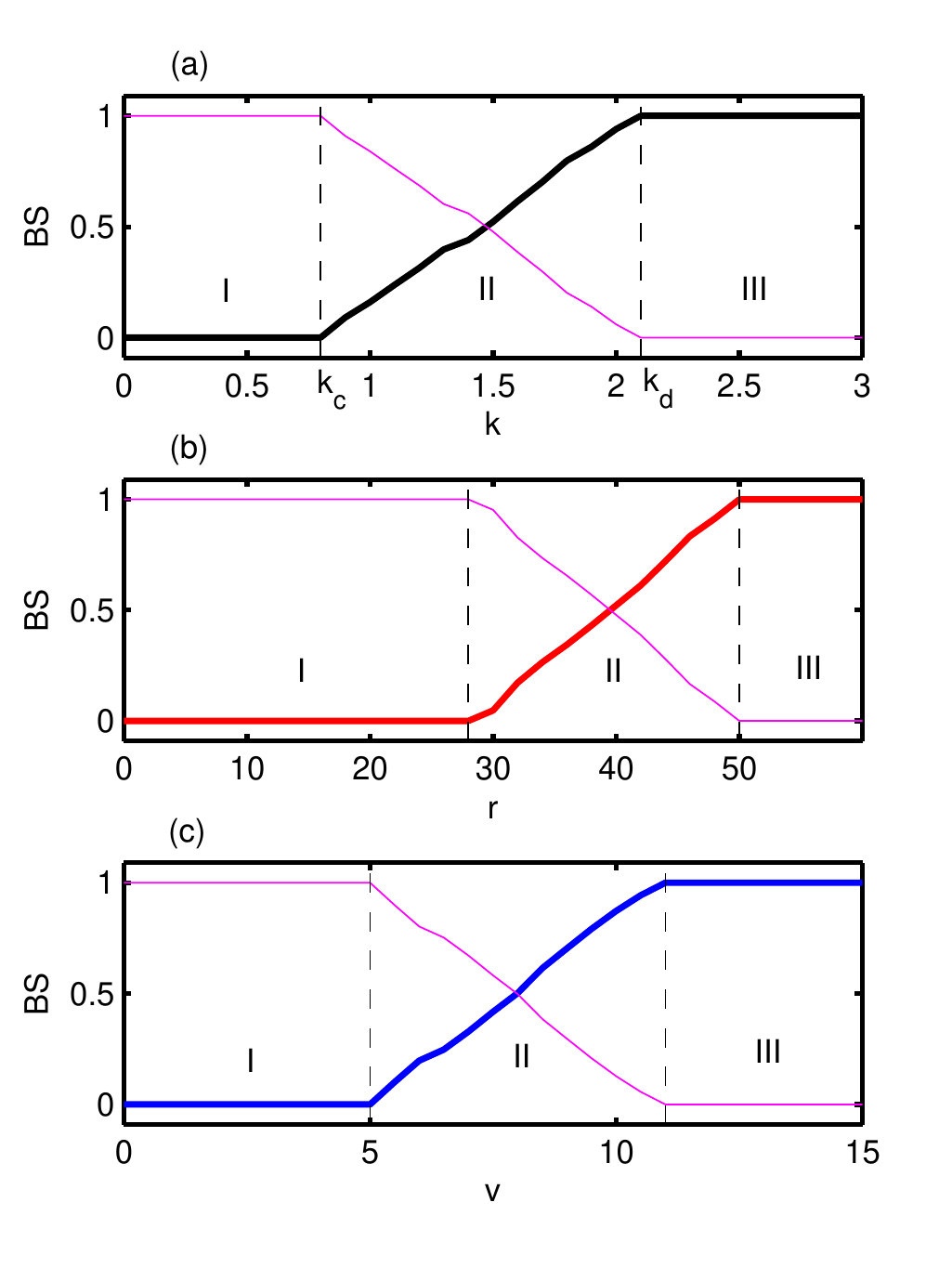}}
		\caption{ Basin stability for moving R\"{o}ssler oscillators: Variation of basin stability BS by changing: $(a)$ the interaction strength $k$ for $v=8$ and $r=35$, $(b)$ vision range $r$ for $v=8$ and $k=2$, $(c)$ speed of movement $v$ for $k=2$ and $r=35$. Here $N=100$. The curve in color magenta stands for the BS of desynchronized state in all the figures whereas the other colors (black, red and blue) represent the BS of synchronized states in the respective figures.}
		\label{bsross}
	\end{figure}
	\par  We start with fixed values of $v=8$ and $r=35$ while vary interaction strength $k$ for $N=100$ moving R\"{o}ssler oscillators' network. As can be noticed in region I=$\{k : 0 \le k < k_c\}$, where $k_c$ is the critical coupling strength obtained analytically in Eq.(12), BS of the synchronized state (black colored curve) remains zero whereas the BS of the desynchronized state (in color magenta) remains unity up to $k=k_c\simeq 0.78$ (Fig.~\ref{bsross}(a)). In this region, the moving oscillators are in desynchronized state whatever be the amount of perturbation from the synchronization manifold. With an increase in the value of $k$ beyond $k=k_c,$  BS of synchrony starts developing in the region II=$\{k : k_c \le k < k_d\}$ where $k_d$ \cite{kd} is the threshold coupling strength for the unit BS.  This indicates that although linear stability predicts $k_c$ as the critical value of $k$ after which the synchrony of the network is stable, this is no more the case when non-small perturbations (random initial conditions) are taken into account. BS of synchronization (desynchronization) becomes unity (zero) in the region III=$\{k: k\ge k_d\}$, which means synchrony of the network is stable here for even numerous choice of initial conditions from the basin volume and hence confirms stability in a global sense. 
	\par Next keeping $v=8$ and $k=2$ fixed, a similar transition from BS zero (unity) to BS unity (zero) for synchrony (desynchronization) is observed for varying vision range $r$ in Fig.~\ref{bsross}(b). Here again, after having zero values in a certain range I, BS of synchrony (in color red) gets non-zero, non-unit values in region II and finally gets unit value in the range III. BS of desynchronization (in magenta color) takes the complementary values in the respective regions. Finally, Fig.~\ref{bsross}(c) characterizes the variation of BS of synchronization (desynchronization) with respect to speed of movement $v$ figured in blue (magenta) colored curve where $k=2$ and $r=35$ are fixed. Here also a similar type of increment (decrement) in the value of BS of synchrony (desynchronization) is observed as before. Similarly the variations of BS for Landau-Stuart oscillators by changing the physical system parameters are discussed in Appendix B.

\section{Conclusion}
 In summary, we have presented a network model of interaction between moving oscillators where each oscillator is moving in three dimensional space. 
	Depending upon the network parameters, one of the most intriguing emergent phenomenon in coupled oscillators system, namely synchronization for networks of spatially moving limit cycle and chaotic oscillators in three dimensional space is observed.  We have investigated the parameter regions for synchronization state by varying the three physical systems' parameters, namely the interaction strength $k$, speed of motion of the moving oscillators $v$ and vision range $r$. Density dependent threshold of coupling strength for synchrony using linear stability analysis has also been determined. Apart from the numerical investigation and linear stability analysis, we probe the notion of basin stability, a nonlinear measure based on volumes of basin of attractions, to investigate how stable the synchronous state is under arbitrary perturbations. 
\par The present study is important in the context of time-varying complex networks where the interactions between elements (species, individuals etc.) that occur in real systems (biological, social etc.) changes with respect to time due to spatial movement of the elements. Since we have considered the collective behaviors of moving oscillators in three-dimensional space, obtained results has a direct application in biological systems such as flashing fireflies \cite{firefl1,firefl2} which are moving in 3D space.                    \\

{\bf APPENDIX A:\\\\
    CONFIRMATION OF RESULTING FORMULA (9) WITH $N=3$\\}
    \par  To confirm the concluding formula given in (9), here we consider $N=3$. In this case, the possible network configurations are based on the following cases:
    	\begin{enumerate}
    		\item[i)] There is no interaction between the nodes. 
    		\item[ii)] Two nodes are interacting with each other but the third one is not interacting with any of these two nodes.
    		\item[iii)] A single node is interacting with the other two nodes but there is no connection between these two nodes.
    		\item[iv)] There exists connection between all the nodes.
    	\end{enumerate}
    	Then the time-average matrix $\bar{G}$ is given by
    	\begin{equation}
    	\begin{array}{lcl}
    	\bar{G}=p_{no} G_{no}+p_{12} G_{12}+p_{13} G_{13}+p_{23} G_{23}+p_1 G_1\\\\
    	~~~~~~~~~~~~~~~+p_2 G_2+p_3 G_3+p_{all} G_{all},
    	\end{array}
    	\end{equation}
    	where $p_{no}$ is the probability that there is no connection between any node. $p_{12}$ is the probability that node $1$ and node $2$ are connected with each other but node $3$ is isolated. $p_{13}$ and $p_{23}$ are similarly defined. $p_1$ is the probability of the case when node $1$ is interacting with both node $2$ and node $3$ but node $2$ and node $3$ are not connected. $p_2$ and $p_3$ have similar meanings. Finally, $p_{all}$ stands for the probability that all the three nodes are interacting with each other. $G_{no}$, $G_{12}$, $G_{13}$, $G_{23}$, $G_1$, $G_2$, $G_3$ and $G_{all}$ are the corresponding Laplacian matrices.
    	\par Combining all the types of interactions (unidirectional and bidirectional), $p_{12} G_{12}$ can be written as
    	\begin{equation}
    	p_{12} G_{12}=p_{\overrightarrow{12}} G_{\overrightarrow{12}}+p_{\overrightarrow{21}} G_{\overrightarrow{21}}+p_{\overline{12}} G_{\overline{12}},
    	\end{equation} 
    	where the probabilities (Laplacian matrices) $p_{\overrightarrow{12}}$ ($G_{\overrightarrow{12}}$) or $p_{\overrightarrow{21}}$ ($G_{\overrightarrow{21}}$) corresponds to the cases when there is a unidirectional connection from node $1$ to node $2$ or from node $2$ to node $1$ and $p_{\overline{12}}$ ($G_{\overline{12}}$) is the probability (Laplacian matrix) that node $1$ and node $2$ are connected bidirectionally.
    	\par Since, $p_{\overrightarrow{12}}=p_{\overrightarrow{21}}=p^*$ (say) and \\
    	$$ G_{\overrightarrow{12}} = \left( \begin{array}{ccccc}
    	0 & 0 & 0\\
    	-1 & 1 & 0 \\
    	0 & 0 & 0\\
    	\end{array} \right)\; \mbox{;}\; G_{\overrightarrow{21}} = \left( \begin{array}{ccccc}
    	1 & -1 & 0\\
    	0 & 0 & 0 \\
    	0 & 0 & 0\\
    	\end{array} \right)$$
    	and $$ G_{\overline{12}} = \left( \begin{array}{ccccc}
    	1 & -1 & 0\\
    	-1 & 1 & 0 \\
    	0 & 0 & 0\\
    	\end{array} \right).$$\\
    	So, from Eq.(15) we obtain  $p_{12} G_{12}=(p^*+p_{\overline{12}}) G_{\overline{12}}.$ Similarly, $p_{13} G_{13}=(p^*+p_{\overline{13}}) G_{\overline{13}}$
    	and $p_{23} G_{23}=(p^*+p_{\overline{23}}) G_{\overline{23}}.$\\
    	\par Again since $p_{\overline{12}}=p_{\overline{13}}=p_{\overline{23}}=pp^*$ (say),\\
    	$p_{12} G_{12}+p_{13} G_{13}+p_{23} G_{23}=(p^*+pp^*) G_g$, where 
    	$$ G_g = \left( \begin{array}{ccccc}
    	2 & -1 & -1\\
    	-1 & 2 & -1 \\
    	-1 & -1 & 2\\
    	\end{array} \right).$$\\
    	In a similar way, taking all the types of interactions into account $p_1 G_1$ becomes
    	\begin{equation}
    	\begin{array}{lcl}
    	p_1 G_1=p_{\overrightarrow{1,23}} G_{\overrightarrow{1,23}}+p_{\overrightarrow{23,1}} G_{\overrightarrow{23,1}}+p_{\overrightarrow{12},\overleftarrow{13}} G_{\overrightarrow{12},\overleftarrow{13}}\\\\
    	~~~~~~~~~~~~~~~+p_{\overleftarrow{12},\overrightarrow{13}} G_{\overleftarrow{12},\overrightarrow{13}}+p_{\overline{1,23}} G_{\overline{1,23}},
    	\end{array}
    	\end{equation}
    	where the probabilities (Laplacian matrices) $p_{\overrightarrow{1,23}}$ ($G_{\overrightarrow{1,23}}$) or $p_{\overrightarrow{23,1}}$ ($G_{\overrightarrow{23,1}}$) are for the cases when there is a unidirectional connection from node $1$ to both node $2$ and $3$ or from both node $2$ and $3$ to node $1$. $p_{\overrightarrow{12},\overleftarrow{13}}$ ($G_{\overrightarrow{12},\overleftarrow{13}}$) or $p_{\overleftarrow{12},\overrightarrow{13}}$ ($G_{\overleftarrow{12},\overrightarrow{13}}$) corresponds to the cases when there is a unidirectional connection from node $1$ to node $2$ and from node $3$ to node $1$ or from node $2$ to node $1$ and from node $1$ to node $3$ and finally $p_{\overline{1,23}}$ ($G_{\overline{1,23}}$) is the probability (Laplacian matrix) that node $1$ connects to both node $2$ and $3$ bidirectionally. But, node $2$ and $3$ are not connected in these cases.\\
    	Since, $p_{\overrightarrow{12},\overleftarrow{13}}=p_{\overleftarrow{12},\overrightarrow{13}}=p_1*$ (say) and \\
    	$$ G_{\overrightarrow{1,23}} = \left( \begin{array}{ccccc}
    	0 & 0 & 0\\
    	-1 & 1 & 0 \\
    	-1 & 0 & 1\\
    	\end{array} \right)\; \mbox{;}\; G_{\overrightarrow{23,1}} = \left( \begin{array}{ccccc}
    	2 & -1 & -1\\
    	0 & 0 & 0 \\
    	0 & 0 & 0\\
    	\end{array} \right)$$
    	$$ G_{\overrightarrow{12},\overleftarrow{13}} = \left( \begin{array}{ccccc}
    	1 & 0 & -1\\
    	-1 & 1 & 0 \\
    	0 & 0 & 0\\
    	\end{array} \right)\; \mbox{;}\; G_{\overleftarrow{12},\overrightarrow{13}} = \left( \begin{array}{ccccc}
    	1 & -1 & 0\\
    	0 & 0 & 0 \\
    	-1 & 0 & 1\\
    	\end{array} \right)$$
    	
    	and $$ G_{\overline{1,23}} = \left( \begin{array}{ccccc}
    	2 & -1 & -1\\
    	-1 & 1 & 0 \\
    	-1 & 0 & 1\\
    	\end{array} \right).$$\\
    	So, Eq.(16) gives $p_1 G_1=p_{\overrightarrow{1,23}} G_{\overrightarrow{1,23}}+p_{\overrightarrow{23,1}} G_{\overrightarrow{23,1}}+p_1^* G_1^*+p_{\overline{1,23}} G_{\overline{1,23}}$
    	with $$ G_1^* = \left( \begin{array}{ccccc}
    	2 & -1 & -1\\
    	-1 & 1 & 0 \\
    	-1 & 0 & 1\\
    	\end{array} \right).$$\\ 
    	Proceeding in the same way, we have\\
    	$p_2 G_2=p_{\overrightarrow{2,13}} G_{\overrightarrow{2,13}}+p_{\overrightarrow{13,2}} G_{\overrightarrow{13,2}}+p_2^* G_2^*+p_{\overline{2,13}} G_{\overline{2,13}}$\\
    	with $$ G_2^* = \left( \begin{array}{ccccc}
    	1 & -1 & 0\\
    	-1 & 2 & -1 \\
    	0 & -1 & 1\\
    	\end{array} \right)$$\\
    	and $p_3 G_3=p_{\overrightarrow{3,12}} G_{\overrightarrow{3,12}}+p_{\overrightarrow{12,3}} G_{\overrightarrow{12,3}}+p_3^* G_3^*+p_{\overline{3,12}} G_{\overline{3,12}}$\\
    	with $$ G_3^* = \left( \begin{array}{ccccc}
    	1 & 0 & -1\\
    	0 & 1 & -1 \\
    	-1 & -1 & 2\\
    	\end{array} \right).$$\\
    	Again since $p_{\overrightarrow{1,23}}=p_{\overrightarrow{2,13}}=p_{\overrightarrow{3,12}}=p_a^*$ (say),\\
    	$p_{\overrightarrow{23,1}}=p_{\overrightarrow{13,2}}=p_{\overrightarrow{12,3}}=p_b^*$ (say),\\
    	$p_{\overline{1,23}}=p_{\overline{2,13}}=p_{\overline{3,12}}=p_c^*$ (say),\\
    	and $p_1^*=p_2^*=p_3^*=p_d^*$ (say) so\\
    	$p_1 G_1+p_2 G_2+p_3 G_3=(p_a^*+p_b^*+2 p_c^*+2 p_d^*) G_g.$\\
    	In this way, taking all possible type of interactions among all the three nodes, we will have $p_{all} G_{all}=p_k^* G_g$, where $p_k^*$ is the sum of all the probabilities corresponding to different type of interactions (considering every possible combination of unidirectional and bidirectional interactions). Thus from Eq.(14), $\bar{G}$ becomes, 
    	$\bar{G}=[p^*+pp^*+p_a^*+p_b^*+2 p_c^*+2 p_d^*+p_k^*] G_g=p_A^* G_g$ (say), where $p_A^*=p^*+pp^*+p_a^*+p_b^*+2 p_c^*+2 p_d^*+p_k^*$ is the probability of activation of a link between three nodes and 
    	$$ G_g = \left( \begin{array}{ccccc}
    	2 & -1 & -1\\
    	-1 & 2 & -1 \\
    	-1 & -1 & 2\\
    	\end{array} \right).$$\\ 
    	Thus  $\bar{G}=\frac{r^3}{L^3} G_g$ where  $p_A^*=\frac{r^3}{L^3}$ and $G_g$ is the zero-row sum Laplacian matrix of order three for global bidirectional interactions.\\\\ 
{\bf APPENDIX B: \\\\
	BASIN STABILITY FOR MOVING LANDAU-STUART OSCILLATORS\\}
\begin{figure}[ht]
	\centerline{
		\includegraphics[scale=0.75]{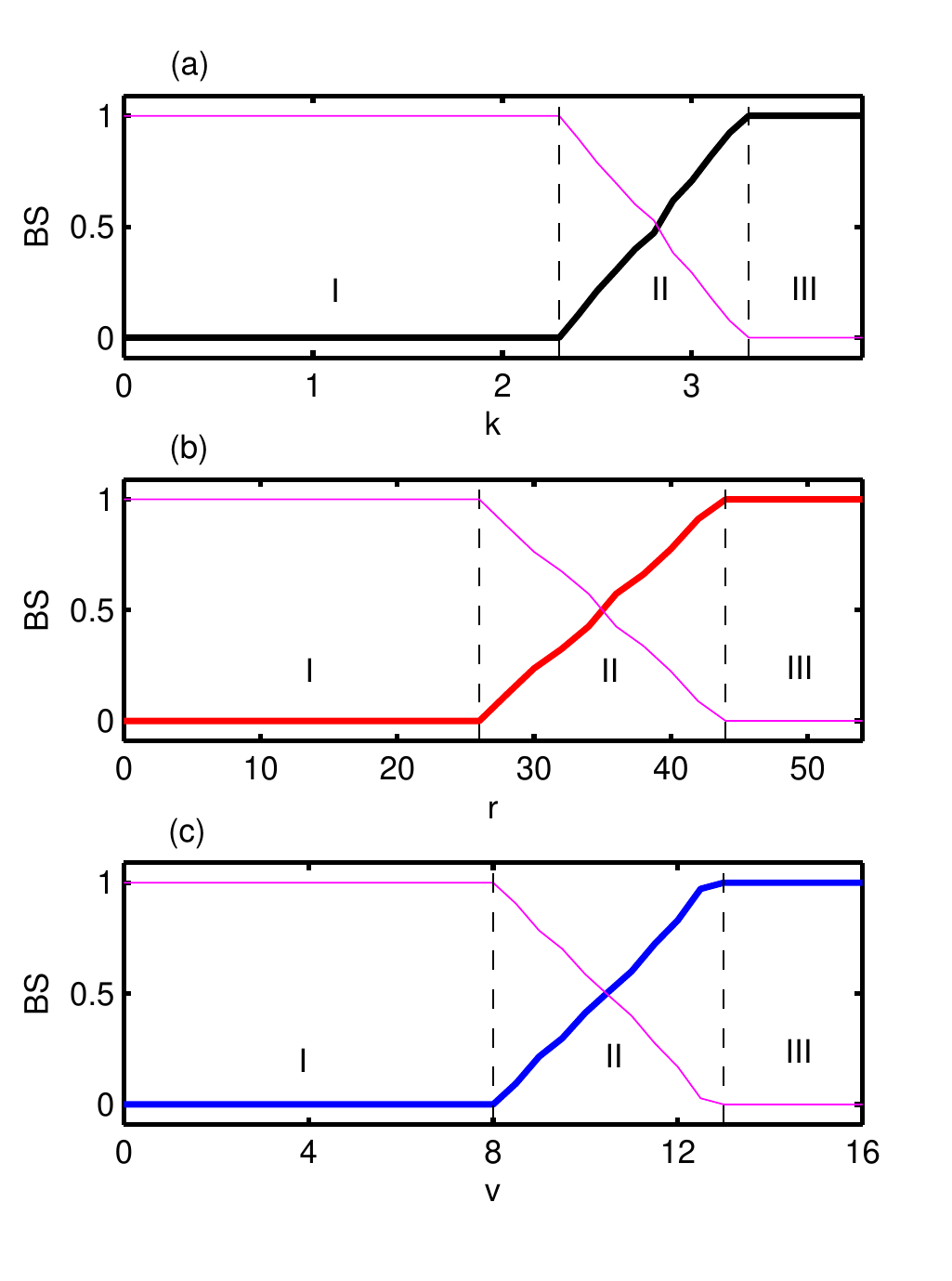}}
	\caption{ Basin stability for moving Landau-Stuart oscillators: Variation of basin stability BS by changing: $(a)$ the interaction strength $k$ for $v=12$ and $r=35$, $(b)$ vision range $r$ for $v=12$ and $k=3.0$, $(c)$ speed of movement $v$ for $k=3.0$ and $r=35$. Here $\omega=10.0$ and $N=100$. The curve in magenta represents the BS of desynchronized state while the other colors (black, red and blue) correspond to the BS of synchronized states in the respective figures. $P=1000$ from the interval $[-1,1]$ is used to calculate BS.  }
	\label{bsls}
\end{figure}
\par  Figure ~\ref{bsls}(a) shows how the BS of the synchronized (desynchronized) state changes with interaction strength $k$ where $v=12$ and $r=35$ are fixed for $N=100$ moving Landau-Stuart oscillators' network. In region I of Fig.~\ref{bsls}(a), BS of the synchronized state (black colored curve) remains zero whereas the BS of the desynchronized state (in color magenta) retains maximum possible value (the unity) upto $k\simeq 2.3$. In region II, BS of synchrony starts increasing as $k$ passes this value, but here this BS is non-unit as well.  BS of synchronization (desynchronization) turns to be unity (zero) in the region III. Next we keep $v=12$ and $k=3.0$ fixed, for which a similar transition from BS zero (unity) to BS unity (zero) for synchrony (desynchronization) is observed for increasing vision range $r$ as in Fig.~\ref{bsls}(b). Here red and magenta colors are used to figure curves representing BS of synchronization and desynchronization respectively. Finally, Fig.~\ref{bsls}(c) depicts the change in BS of synchronization (desynchronization) with respect to speed of movement $v$ where $k=3.0$ and $r=35$ are kept fixed. \\
\\
\par {\bf Acknowledgments:}
Authors thank the anonymous referees for valuable suggestions to improve the manuscript in the present form.   D.G. was supported by SERB-DST (Department of Science and Technology), Government of India (Project no. EMR/2016/001039).


\begin{thebibliography}{25}
	
	
	\bibitem{syper}  I. Blekman, Synchronization in Science and Technology, Nauka, Moscow, 1981 (in Russian); ASME Press, New
	York, 1988 (in English)
	
	\bibitem{sycha}  S. Boccaletti, J. Kurths, G. Osipov, D. L. Valladares, C. S. Zhou, {Phys. Rep. {\bf 366},  1-101 (2002).}
	
	
	
	\bibitem{pts}  M. L. Sachtjen, B. A. Carreras, and V. E. Lynch, {Phys. Rev. E {\bf 61},  4877 (2000).}
	
	\bibitem{cons}  R. Olfati-Saber, J. A. Fax, and R. M. Murray, {Proc. IEEE {\bf 95},  215 (2007).}
	
	\bibitem{mcom}   J.-P. Onnela, J. Saramaki, J. Hyvonen, G. Szabo, D. Lazer, K. Kaski, J. Kertesz, and A. L. Barabasi, {Proc. Natl. Acad. Sci. USA {\bf 104},  7332 (2007).}
	
	\bibitem{mcom2} Y. Wu, C. Zhou, J. Xiao, J. Kurths, and H. J. Schellnhuber, {Proc. Natl. Acad. Sci. U. S. A. {\bf 107}, 18803 (2010).}
	
	\bibitem{mcom3} J. L. Iribarren and E. Moro, {Phys. Rev. Lett. {\bf 103}, 038702 (2009).}
	
	\bibitem{mcom4} R. Guimer\`{a}, L. Danon, A. D\'{\i}az-Guilera, F. Giralt, and A. Arenas, {Phys. Rev. E {\bf 68}, 065103 (2003).}
	
	\bibitem{fbn}  M. Valencia, J. Martinerie, S. Dupont, and M. Chavez, {Phys. Rev. E {\bf 77},  050905(R) (2008).}
	
	\bibitem{belykh}  I. V. Belykh, V. N. Belykh, and M. Hasler, {Physica D {\bf 195},  188 (2004).}
	
	\bibitem{sinha}  V. Kohar, P. Ji, A. Choudhary, S. Sinha and J. Kurths, {Phys. Rev. E {\bf 90},  022812 (2014).}
	
	\bibitem{sychen1}  J. L\"{u}, G. Chen, {IEEE Trans. Autom. Control {\bf 50},  841 (2005).}
	
	\bibitem{sychen2}  M. Chen, {Phys. Rev. E {\bf 76},  016104 (2007).}
	
	\bibitem{danas}  S. K. Bhowmick, R. E. Amritkar, and S. K. Dana, {Chaos {\bf 22},  023105 (2012).}
	
	
	\bibitem{grani}  J. Buhl, D. J. T. Sumpter, I. D. Couzin, J. J. Hale, E. Despland, E. R. Miller, and S. J. Simpson, {Science {\bf 312},  1402 (2006).}
	
	\bibitem{robot}  A. Buscarino, L. Fortuna, M. Frasca, and A. Rizzo, {Chaos {\bf 16},  015116 (2006).}
	
	\bibitem{vehicle}  H. G. Tanner, A. Jadbabaie, and G. J. Pappas, in \emph {Proceedings, 42nd IEEE Conference on Decision and Control}, pp. 2016-2021 (2003).
	
	\bibitem{chemo}  D. Tanaka, {Phys. Rev. Lett. {\bf 99},  134103 (2007).}
	
	\bibitem{sensor}  F. Sivrikaya, and B. Yener, {IEEE Network {\bf 18},  45 (2004).}
	
	\bibitem{adhoc}  K. R\"{o}mer, in \emph{Proceedings of the 2nd ACM International Symposium on Mobile ad hoc Networking \& Computing}
	(ACM, NY, USA, 2001), pp. 173-182.
	
	\bibitem{movkur}  N. Fujiwara, J. Kurths, A. D´ıaz-Guilera, {Phys. Rev. E {\bf 83},  025101(R) (2011).}
	
	
	\bibitem{movch}  M. Frasca, A. Buscarino, A. Rizzo, L. Fortuna and S. Boccaletti, {Phys. Rev. Lett. {\bf 100},  044102 (2008).}
	
	\bibitem{spect}  N. Fujiwara, J. Kurths, A. D\'{i}az-Guilera, {AIP Conf.
		Proc. {\bf 1389},  1015 (2011).}
	
	
	\bibitem{infect}  M. Frasca, A. Buscarino, A. Rizzo, L. Fortuna and S. Boccaletti, {Phys. Rev. E {\bf 74},  036110 (2006).}
	
	\bibitem{punit}  L. Janagal, P. Parmananda, {Phys. Rev. E {\bf 86},  056213 (2012).}
	
	
	\bibitem{if1}  L. Prignano, O. Sagarra, A. D\'{i}az-Guilera, {Phys. Rev. Lett. {\bf 110},  114101 (2013).}
	
	\bibitem{if2}  L. Prignano, O. Sagarra, P. M. Gleiser, A. D\'{i}az-Guilera, {Int. J. Bifurcat. Chaos {\bf 22},  1250179 (2012).}
	
	 \bibitem{rev1}   R. Gro\ss mann, F. Peruani, and M. B\"{a}r, {Phys. Rev. E {\bf 93},  040102(R) (2016).}

    \bibitem{rev2}  D. Levis, I. Pagonabarraga, and A. D\'{i}az-Guilera, {Phys. Rev. X {\bf 7},  011028 (2017).}
	
	\bibitem{rev3}  B. Liebchen, M. E. Cates and D. Marenduzzo, {Soft Matter {\bf 12},  7259 (2016).}
	
	\bibitem{wang}  L. Wang, H. Shi, and Y-X. Sun, {Phys. Rev. E {\bf 82},  046222 (2010).}
	
	\bibitem{pori}  M. Porifiri, D. J. Stilwell, E. M. Bollt and J. D. Skufca, {Physica D {\bf 224},  102 (2006).}
	
	\bibitem{chaos2016}  N. Fujiwara, J. Kurths, and A. D\'{\i}az-Guilera, {Chaos {\bf 26},  094824 (2016).}
	
	\bibitem{kim}  B. Kim, Y. Do, and Y-C. Lai, {Phys. Rev. E {\bf 88},  042818 (2013).}
	
	
	\bibitem{firefl1}  R. E. Mirollo and S. H. Strogatz, {SIAM (Soc.	Ind. Appl. Math.) J. Appl. Math. {\bf 50},  1645 (1990).}
	
	\bibitem{firefl2}  G. M. Ram\'{i}rez-Avila, J.-L. Deneubourg, J.-L. Guisset, N. Wessel and J. Kurths, {Euro. Phys. Lett. {\bf 94},  60007 (2011).}
	
	\bibitem{initc} While calculating all the order parameters (throughout the paper, in Figs. 2 , 4 and 5),  we have taken the initial conditions very  near to the synchronization manifold  $ X_i=X_j, \forall i,j$ so that the order parameter  $\phi$  reaches 1 (the unity) exactly when $ k=k_c $ i.e., at the earliest possible. 
	
	\bibitem{stilwell}  D. J. Stilwell, E. M. Bollt, and D. G. Roberson, {SIAM J. Appl. Dyn. Syst. {\bf 5},  140 (2006).}
	
	\bibitem{msf1} L. M. Pecora and T. L. Carroll, {Phys. Rev. Lett. {\bf 80},  2109 (1998).}
	
	\bibitem{msf2}  M. Barahona and L. M. Pecora, {Phys. Rev. Lett. {\bf 89},  054101 (2002).}

   \bibitem{msf3}  L. Huang, Q. Chen, Y-C. Lai, and L. M. Pecora, {Phys. Rev. E. {\bf 80},  036204 (2009).}
    	
    \bibitem{msf4}  J. Sun, E. M. Bollt and T. Nishikawa, {Euro. Phys. Lett. {\bf 85},  60011 (2009).}
    	
	\bibitem{msf5}  A. E. Hramov, A. E. Khramova, A. A. Koronovskii, and S. Boccaletti, {Int. J. Bifurcat. Chaos {\bf 18},  845 (2008).}
	
	\bibitem{quo1}  A. Camili and B. L. Bassler, {Science {\bf 311},  1113 (2006).}
	
	\bibitem{quo2}  A. F. Taylor, M. R. Tinsley, F. Wang, Z. Huang, K. Showalter, {Science {\bf 323},  614 (2009).}
	\bibitem{quo3}  S. Majhi, M. Perc, and D. Ghosh, Sci. Rep. {\bf 6}, 39033 (2016).
	\bibitem{quo4}  C. D. Nadell, J. Xavier, S. A. Levin, and K. R. Foster, {PLoS Comput. Biol. {\bf 6},  e14 (2008).}	
	
	\bibitem{jost}  J. Jost and M. P. Joy, {Phys. Rev. E {\bf 65},  016201 (2001).}
	
	\bibitem{bs}  P. J. Menck, J. Heitzig, N. Marwan and J. Kurths, {Nature Phys. {\bf 9},  89 (2013).}
	
	\bibitem{tol} In this context, $\phi$ satisfying $0.995 \le \phi \le 1.0$ has been taken into consideration that corresponds to synchronizability of the network. 
	
	\bibitem{kd} The value of $k_d$ may vary depending on the choice of $P$. 
	
	
	
\end{thebibliography}
\end{document}